\documentstyle[11pt,epsf,shst2] {article}
\begin{document}

\title{H$_0$ from HST}

\author{G.A.~Tammann, L.~Labhardt \& M.~Federspiel}
\affil{Astronomisches Institut der Universit\"at Basel, Venusstr.~7,
       CH-4102 Binningen, Switzerland}

\author{A.~Sandage}
\affil{The Observatories of the Carnegie Institution, 813 Santa Barbara
       Street, Pasadena, CA 91101, USA}

\author{A.~Saha, F.D.~Macchetto \& N.~Panagia}
\affil{Space Telescope Science Institute
       3700 San Martin Drive, Baltimore, MD 21218, USA}
\begin{abstract}
{\it HST} has so far provided Cepheid distances to nine galaxies. Although
not sufficient yet to determine the distance of the {\it extended} Virgo
cluster, they are decisive for the distance scale in two ways. (1) Seven of
the galaxies contribute to a much improved calibration of the Tully-Fisher
relation.  Applying this to a {\it complete} sample of Virgo spirals one
obtains a cluster distance of $(m-M)=31.79\pm 0.09$. Other distance
indicators support this value. The adopted linear distance of $22.0\pm
0.8$~Mpc combined with the cluster velocity of $1178\pm 32~{\rm km\,s^{-1}}$
(in the CMB frame) gives $H_{0}=54\pm 2$ (internal error). (2) Six of the
galaxies have been the site of seven SNe Ia with well observed maxima. Their
resulting calibration in absolute magnitudes gives $M_{\rm B}({\rm
max})=-19.53\pm 0.07$ and $M_{\rm V}({\rm max})=-19.49\pm 0.07$ with
negligible {\it intrinsic} scatter.  If this calibration is used to
determine the distances of {\it all} distant SNe Ia with known maxima and
with $1100 < v < 30\,000\ {\rm km\,s^{-1}}$, $H_{0}$ becomes $56\pm 3$
(internal error). Systematic errors tend to make this an upper limit; in
particular the case $H_{0}\ge 70$ can be excluded. -- The conclusion is that
the large-scale value of the Hubble constant is $H_{0}=55\pm 10$ (external
error).
\end{abstract}
\section{Introduction}
The influence of {\it HST} on the determination of $H_{0}$ is already
enormous and it can only grow. In Table~1 a compilation is given of
distance determinations with {\it HST} having some bearing on $H_{0}$.  The
values scatter between 55 and 80 and the formal mean is $H_{0}=61\pm 3$.
This, however, is not the best value, because some values shown are
mutually incompatible.

There are now two self-consistent routes to $H_{0}$ to which {\it HST} has
heavily contributed. The first route via the Virgo cluster is
described in Section 2, the second route using supernovae of type Ia
(SNe Ia) is discussed in Section 3.

\begin{table}[thb]
\caption{$H_{0}$ determinations from {\it HST}}
\begin{center} \scriptsize
\begin{tabular}{llr}
\tableline
\noalign{\smallskip}
\multicolumn{2}{l}{I.~Cepheids} & \\
\noalign{\smallskip}
a) in M101 & Kelson (1995) & \\
           & confirms Sandage \& Tammann (1974), & \\
           & from which follows                  & $55\pm \phantom{1}9$ \\
\noalign{\smallskip}
b) in Leo Group & Tanvir et al.~(1995)           & $69\pm \phantom{1}8$ \\
                & present paper                  & $57\pm \phantom{1}6$ \\
\noalign{\smallskip}
c) in Virgo Cluster & & \\
   \quad NGC 4321 & Freedman et al.~(1994)     & $80\pm 17$ \\
   \quad NGC 4639 & Sandage et al.~(1996)      & $47\pm 10$ \\
\noalign{\smallskip}
\tableline
\noalign{\smallskip}
\multicolumn{3}{l}{II.~Tully-Fischer distance of Virgo cluster
using 11 Calibrators}  \\ 
\multicolumn{3}{l}{\quad with Cepheid Distances
(7 of which from HST)} \\
\noalign{\smallskip}
                   & Federspiel et al.~(1996) & $52\pm \phantom{1}6$ \\
\noalign{\smallskip}
\tableline
\noalign{\smallskip}
\multicolumn{3}{l}{III.~SNe Ia calibrated through Cepheids} \\
\noalign{\smallskip}
 SN 1937C & Saha et al.~(1994)           & $52\pm \phantom{1}9$ \\
 SN 1972E & Hamuy et al.~(1995)           & $65\pm 10$ \\
 SN 1972E & Riess et al.~(1995)           & $67\pm \phantom{1}7$ \\
 SN 1895B & Schaefer (1995)               & $51\pm \phantom{1}7$ \\
 6 SNe Ia & Branch et al.~(1996)           & $57\pm \phantom{1}7$ \\
 7 SNe Ia & Sandage et al.~(1996)           & $58\pm \phantom{1}4$ \\
\noalign{\smallskip}
\tableline
\noalign{\smallskip}
\multicolumn{3}{l}{IV.~Globular Clusters} \\
 in M87 & Whitmore et al.~(1995) & $78\pm 11$ \\
         & Sandage \& Tammann (1996) & $62\pm \phantom{1}9$ \\
 in Coma & Baum et al.~(1995) & $<65$ \\ 
\noalign{\smallskip}
\tableline \tableline
\end{tabular}
\end{center}
\end{table}
Much of what follows depends on Cepheid distances. A word on the
reliability of their P-L relation is therefore in place. The P-L
relation of Madore \& Freedman (1991), adopted in the following, {\it
assumes} an LMC modulus of 18.50. Actual confirmation of this value to
better than 0.10~mag comes from the P-L relation calibrated by
Galactic Cepheids (Sandage \& Tammann 1968; Feast \& Walker 1987) and
independently from the ring of SN~1987A (Panagia et al.~1991; Gould
1994); further support is given by RR~Lyr stars and other distance
determinations (cf. Tammann 1996). The zero point of the adopted P-L
relation is therefore secure to $<5\%$ in linear distance.  The slope
of the relation is well determined by the LMC Cepheids; it is of less
importance as long as the Cepheids under consideration cover a
sufficient period range, which is also needed to avoid selection
effects (Sandage 1988). Metallicity effects are believed to be small
(Freedman \& Madore 1990; Chiosi et al.~1993).
\section{The Distance of the Virgo Cluster}
\subsection{Cepheids}
When the first reliable Cepheid distance of a Virgo galaxy (NGC~4321)
became available from {\it HST} (Freedman et al.~1994) it was
precipitately hailed {\it the} Virgo cluster distance (Mould et
al.~1995; Kennicutt et al.~1995), although the value of 17~Mpc was
suspiciously low. The next two Virgo galaxies, NGC~4536 (Saha et
al.~1996a) and NGC~4496A (Saha et al.~1996b), again had very low
distances. Only the fourth galaxy, NGC~4639, revealed the {\it
important depth} of the cluster (note for comparison: the spiral
members span $\sim 15\deg$ in the sky!). Its distance is 25~Mpc
(Sandage et al.~1996) and yet it is a bona fide cluster member; with a
{\it small} velocity of 800~km\,s$^{-1}$ it cannot be assigned to the
background.

It is no accident that the first three Virgo spirals with Cepheid
distances lie on the near side of the cluster. They were selected from
Sandage \& Bedke's (1988) atlas of galaxies most suited for {\it HST};
they were thus biased to begin with. NGC~4639 looks much more
difficult and would not have been selected had it not produced the
archetypal SN 1990N.

It is now clear that it will take at least a dozen Cepheid distances
of a randomly selected sample of Virgo members to obtain a meaningful
mean cluster distance.

Meanwhile Tanvir et al.~(1995) have suggested to step up their 
Cepheid-based Leo group distance of $(m-M)=30.32\pm 0.16$ out to the
Virgo cluster by means of {\it relative} distance determinations. The 
best available relative distances are compiled in Table~2. Adding the
mean difference of $\Delta (m-M)=1.25\pm 0.13$ to the above distance 
modulus gives a Virgo cluster modulus of $(m-M)_{\rm Virgo}=31.57\pm 0.21$.
For brevity we will refer to this value in the following as the 
\lq\lq Cepheid distance of the Virgo cluster\rq\rq .

\begin{table}
\noindent \caption{The distance modulus difference between the Leo group
and the Virgo cluster}
\begin{center} \scriptsize
\begin{tabular}{lll}
\noalign{\medskip}
\tableline
\noalign{\smallskip}
 Method & $\Delta (m-M)_{\rm Virgo - Leo}$ & Source \\
\noalign{\smallskip}
\tableline
\noalign{\smallskip}
 Tully-Fischer & $1.35\pm 0.20$ & Federspiel et al.~(1996) \\
 Globular clusters & $1.47\pm 0.42$ & Harris (1990) \\
 D$_{\rm n}-\sigma$ & $0.97\pm 0.29$ & Faber et al.~(1989) \\
 Planetary nebulae & $1.15\pm 0.30$ & Bottinelli et al.~(1991) \\
 Velocities & $1.30\pm 0.30$ & Kraan-Korteweg (1986) \\
\noalign{\smallskip}
\tableline
\noalign{\medskip}
 mean: & $1.25\pm 0.13$ & \\
\noalign{\smallskip}
\tableline \tableline
\end{tabular}
\end{center}
\end{table}

\subsection{The Tully-Fisher Relation}
There are now 11 spiral galaxies with Cepheid distances (seven of which
come from {\it HST}) which are useful for the calibration of the relation 
between absolute magnitude and $\log w$ ($w=$ inclination-corrected 21cm-line
width). Two close companions of M101 bring the number of useful calibrators
to 13 (cf.~Table~3). Two galaxies (M101 and M100) are less inclined than 
the frequently adopted limit of $i=45\deg$; yet their inclinations are
so well defined by mapping their velocity field that they are still useful 
as calibrators.

\begin{table}
\caption{Galaxies with Cepheid distances for the
calibration of the Tully-Fisher relation}
\begin{center} \scriptsize
\begin{tabular}{lcllccl}
\tableline
\noalign{\smallskip}
 Name & Hubble- &  $(m-M)$ & Source & $M_{\rm B}$  & 
 $i_{\rm RC3}$ & $\log w$ \\ 
      & type & \quad mag & & mag & $^{\circ}$ & \\
 (1) & (2) & \quad (3) & (4) & (5) & (6) & (7) \\
\noalign{\smallskip}
\tableline
\noalign{\smallskip}
 N224 (M31) & 3 & 24.44 & Madore \& Freedman (1991)& -21.10& 78 & 2.739 \\ 
 N300 & 5 & 26.67 & Madore \& Freedman (1991) & -18.14 & 44 & 2.344 \\
 N598 (M33) & 5 & 24.63 & Madore \& Freedman (1991) & -18.89 & 55 & 2.373 \\
 N2403 & 5 & 27.51 & Tammann \& Sandage (1968) & -19.19 & 62  & 2.484 \\
       &   &       & Madore \& Freedman (1991) &        &     &       \\
 N3031 (M81) & 3 & 27.80$^{\ast}$ & Freedman \& Madore (1994) & -20.49 & 65 & 
   2.697 \\
 N3368 (M96) & 2 & 30.32$^{\ast}$ & Tanvir et al.~(1995) & -20.55 & 50 & 
   2.656 \\
 N4321 (M100) & 5 & 31.16$^{\ast}$ &Freedman et al.~(1994) & -21.25 & 36 & 
   2.786 \\
 N4496A & 5 & 31.13$^{\ast}$ & Saha et al.~(1996b)& -19.46 & 43 & 2.428 \\
 N4536 & 4 & 31.11$^{\ast}$ & Saha et al.~(1996a) & -20.64 & 66 & 2.546 \\
 N4639 & 3 & 32.00$^{\ast}$ & Sandage et al.~(1996)& -20.10 & 50 & 2.626 \\
 N5204 & 7 & 29.30& like M101 & -17.93 & 57 & 2.146 \\
 N5457 (M101) & 5 & 29.30$^{\ast}$ & Kelson (1995) & -21.11 & 27 & 2.588 \\
 N5585 & 7 & 29.30 & like M101 & -18.26 & 52 & 2.290 \\
\noalign{\smallskip}
\tableline\tableline
\noalign{\smallskip}
\multicolumn{3}{l}{$^{\ast}$ Cepheid distance from {\it HST}} & & & & \\
\end{tabular}
\end{center}
\end{table}

The date in Table~3 yield the following calibration of the TF relation
\begin{equation}
M_B=-6.39 \log w - 3.80 \ \ \ \ (\sigma = 0.44)
\end{equation}
(Federspiel et al.~1996), where the slope is taken from the Virgo cluster.

An {\it objective} and {\it complete} sample of Virgo spirals is
defined by the 48 non-peculiar galaxies of type Sab--Sdm from the
Virgo Cluster Catalog (Binggeli et al.~1985) with $i \ge 45\deg$ and
lying within the isopleths of substructures A and B (see Binggeli et
al.~1993) or, without changing the result, within the X-ray contour of
the cluster (B\"ohringer et al.~1994). This sample together with
equation (1) gives $(m-M)_{\rm Virgo}=31.79\pm 0.15$. The use of infrared
instead of $B$ magnitudes does not bring an advantage (Schr\"oder
1995), nor does the application of the {\it inverse} TF relation. For
the robustness of the result against variations of the input
parameters the reader is referred to Federspiel et al.~(1996).

\subsection{Other Distances to the Virgo Cluster}
The peak of the luminosity function (LF) of {\it globular clusters}
(GC) has frequently been used as a standard candle. A modern
calibration of the GCs in the Galaxy and in M31 combined with a
compilation of published GCLFs in five Virgo ellipticals has led to a
Virgo modulus of $(m-M)_{\rm Virgo}=31.75\pm 0.11$ (Sandage \& Tammann
1995). Meanwhile Whitmore et al.~(1995) found a very bright peak
magnitude in $V$ and $I$ for NGC~4486, which is well determined with
{\it HST} and which corresponds, with our precepts, to a modulus of
$31.41\pm 0.28$ (Sandage \& Tammann 1996). However, the GCs in
NGC~4486 have a bimodal color distribution which is suggestive of age
differences and possible merger effects (Fritze-von Alvensleben 1995;
Elson \& Santiago 1996). Turning a blind eye to this problem and
averaging over all available GCLFs in Virgo we obtain
$(m-M)_{\rm Virgo}=31.67\pm 0.15$. We are aware that the method may still
face considerable problems.

The {\it $D_n-\sigma$ method}, normally applied to ellipticals, was extended
to the bulges of S0 and spiral galaxies by Dressler (1987). Using the bulges
of the Galaxy, M31, and M81 as local calibrators, one obtains $(m-M)_{\rm
Virgo} =31.85\pm 0.19$ (Tammann 1988).

{\it Novae} are potentially powerful distance indicators through their
luminosity-decline rate relation. Using the Galactic calibration of
Cohen (1985), Capaccioli et al.~(1989) have found the {\it }apparent
distance modulus of M31 to be $(m-M)_{\rm AB}=24.58\pm 0.20$
(i.e. somewhat less than indicated by Cepheids). From six novae in
three Virgo ellipticals Pritchet \& van den Bergh (1987) concluded
that the cluster is more distant by $7.0\pm 0.4$~mag than the apparent
modulus of M31, implying $(m-M)_{\rm Virgo}=31.58\pm 0.45$ (zero
absorption is adopted for the Virgo cluster, see Section 2.6). The
result carries still small weight, but is interesting because it is
based on novae exclusively. {\it HST} observations, although
time-consuming, of novae in the Virgo cluster could much improve this
{\it independent} result.

Theoretical {\it models of SNe Ia} by various authors converge towards
$M_{\rm B}=-19.45\pm 0.15$ for ``Branch normal'' objects (Branch 1996;
H\"oflich \& Khokhlov 1996; Ruiz-Lapuente 1996). It is true that fainter,
nearby SNe Ia are known, but being red and spectroscopically peculiar they
can easily be singled out, and they do not contaminate distant,
luminosity-selected samples of SNe Ia (cf.~Section 3). Eight SNe Ia which
have occurred in the Virgo cluster have $\langle m_{\rm B}({\rm max})\rangle
=12.10\pm 0.15$~mag. This value combined with the theoretical calibration
gives $(m-M)_{\rm Virgo}=31.55\pm 0.25$. -- Had we used instead the
empirical calibration of Table~5 below, the Virgo modulus would have become
larger by 0.08~mag. We refrain from using this value because the routes
towards $H_{0}$ in Sections 2 and 3 are to be kept strictly apart.

\subsection{Suspicious Distance Indicators} 
The assumption that the LF of the {\it shells of planetary nebulae}
(PN) in the light of the $5007~$\AA~line had a universal cutoff at
$M_{5007}=-4.48$~mag has led to a Virgo modulus significantly lower
than obtained from the six methods discussed above (Jacoby et
al.~1990). Yet it was pointed out that the cutoff magnitude depends
on the sample size (i.e.~the absolute magnitude of the parent galaxy;
Bottinelli et al.~1991; Tammann 1993). Numerically simulated LFs of
the shell luminosities confirm indeed the dependence on sample size
{\it and} population age (M\'endez et al.~1993). As a consequence the
published PN distances deviate systematically from the Cepheid
distances. The deviations increase with distance from M81 (Jacoby 
et al.~1989), NGC~5253 (Jacoby \& Ciardullo 1993), and the Leo 
group (Ciardullo et al.~1989) to reach at the Virgo cluster
(Jacoby et al.~1990) 0.74~mag ! -- A new method to derive PN
distances, allowing for sample size and other effects, has been
proposed by Soffner et al.~(1995); the first result for the nearby
galaxy NGC~300 is encouraging.

{\it Surface brightness fluctuations} (SBF) have also been proposed as
distance indicators (Tonry \& Schneider 1988). The first ``test'' has
remained rather unconvincing, spreading the {\it elliptical} Virgo
cluster members over an interval of 12 to 24 Mpc (Tonry et al.~1990);
this interval was interpreted as real although early-type galaxies are
known to be concentrated in the {\it cores} of galaxy clusters. 
Moreover the individual distances correlate with the Mg$_{2}$ index
(Lorenz et al.~1993). Finally we note that the SBF distances of
NGC~5253 (Phillips et al.~1992), the Leo group (Tonry 1991), and the
Virgo cluster (Tonry 1991) are smaller than Cepheid distances by as
much as 0.97, 0.48, and 0.56~mag, respectively.

\noindent For the said reasons we use neither the PN nor the SBF distances.

\subsection{The Structure and Velocity of the Virgo Cluster} A census of the
Virgo cluster containing almost 2000 certain and possible members (Binggeli
et al.~1985) reveals a complex structure with two main subclusters A and B
and additional concentrations some of which, particularly in the
south-western part, are more distant. To obtain a genuine cluster sample we
restrict the sample to the 364 galaxies with known redshifts lying within
the outer isopleths of subclusters A and B (Binggeli et al.~1993; the
individual galaxies are listed there). Their mean velocity is $\langle
v_0\rangle = 918\pm 35\ {\rm km\,s^{-1}}$ (with respect to the Local Group
centroid).  If one considers instead the 361 galaxies within the very
similar X-ray contour of the cluster (B\"ohringer et al.~1994) the mean
velocity becomes $983\pm 39 \ {\rm km\,s^{-1}}$. Taking all 385 galaxies
with redshifts in the Virgo survey area, excluding only background objects,
we find $\langle v_0\rangle = 937\pm 35\ {\rm km\,s^{-1}}$.  From this we
adopt a best cluster velocity of $\langle v_0\rangle = 950\pm 30\ {\rm
km\,s^{-1}}$.  This result supersedes an early value of $\langle v_0\rangle
=1073\pm 50\ {\rm km\,s^{-1}}$ (Huchra 1988) which was based on only 250
galaxies and a less well defined area.

To obtain the {\it cosmic} recession velocity of the Virgo cluster the
observed value must still be corrected for the deceleration of the Local
Group. We adopt $v_{\rm infall}=220\pm 50\ {\rm km\,s^{-1}}$ (cf.~Tammann
1996) and find for Virgo $v_{\rm cosmic}=1170\pm 61\ {\rm km\,s^{-1}}$

Yet we prefer a very similar value the rational of which, however, is quite
different. Many authors step up the Virgo distance out to the Coma cluster
using the {\it relative} distance modulus between Virgo and Coma, and find
$H_{0}$ at the distance of Coma. One can repeat that with any cluster whose
distance relative to Virgo is known. In fact there are at least 14 clusters
with rather good relative distances and velocities $4000<v^{\rm
CMB}<11\,000\ {\rm km\,s^{-1}}$ ($v^{\rm CMB}$ is the velocity in the frame
of the microwave background). The best cosmic value of $H_{0}$ is then the
all-sky mean over 14 different $H_{0}$ determinations. But more elegant is
the reverse method: the relative distances are used to scale down the
velocities of the 14 clusters and to predict a mean Virgo cluster velocity,
i.e.~the $v^{\rm CMB}_{\rm Virgo}$ velocity which the cluster would have in
the absence of all local deviations from an ideal expansion field.  The
result of this procedure is $v^{\rm CMB}_{\rm Virgo}=1178\pm 32\ {\rm
km\,s^{-1}}$ (Sandage \& Tammann 1990, Jerjen \& Tammann 1993; Jerjen 1995
for a more rigid error determination).

\subsection{The Mean Virgo Cluster Distance and H$_{0}$} 
The six independent distance determinations of the Virgo cluster in Sections
2.1 -- 2.3 are repeated in Table~4.

All distance moduli are taken to be true values, i.e.~zero absorption
is assumed towards the Virgo cluster. If the $B$-absorption implied by
Burstein \& Heiles (1984) is applied individually to all galaxies used
for the distance determinations, the modulus becomes lower by only
0.06 mag. Even this almost negligible amount may be an overestimation
as discussed by Sandage \& Tammann (1996).

\begin{table}
\noindent \caption{Distance moduli of the Virgo cluster}
\begin{center} \scriptsize
\begin{tabular}{ll}
\noalign{\smallskip}
\tableline
\noalign{\smallskip}
 Method & $(m-M)_{\rm Virgo}$ \\
\noalign{\smallskip}
\tableline
\noalign{\smallskip}
 Cepheids (via Leo) & $31.57\pm 0.21$ \\
 Tully-Fischer & $31.79\pm 0.15$ \\
 Globular clusters & $31.67\pm 0.15$ \\
 D$_{\rm n}-\sigma$ & $31.85\pm 0.19$ \\
 Novae & $31.58\pm 0.45$ \\
 Theor.~Supernovae & $31.55\pm 0.25$ \\
\noalign{\smallskip}
\tableline
\noalign{\smallskip}
 unweighted mean: & $31.67\pm 0.05$ \\
 weighted mean:   & $31.71\pm 0.08$ \\
 mean linear distance: & $22.0\pm 0.8$ Mpc \\
\noalign{\smallskip}
\tableline \tableline
\end{tabular}
\end{center}
\end{table}

If the adopted mean cluster distance of $22.0\pm 0.8$ Mpc is combined with
$v^{\rm CMB}_{\rm Virgo}=1178\pm 32\ {\rm km\,s^{-1}}$ one obtains
\begin{equation}
  H_{0}=54\pm 2\ {\rm (internal\ error).}
\end{equation}
\section{H$_0$ from SNe Ia}
An {\it HST} program has been mounted to determine the large-scale value
of $H_{0}$. The aim is to derive Cepheid distances (in $V$ and $I$ to
control absorption effects) of up to ten galaxies which have produced
well observed SNe Ia. So far we have calibrated the peak luminosity of
six SNe Ia. A seventh object has become available through Tanvir's et
al.~(1995) Cepheid distance of the Leo group. (It can be assumed that
the member galaxies of this compact group lie practically at the same
distance.) The resulting absolute magnitudes of the seven SNe Ia are
shown in Table~5. Detailed discussions of the input parameters are
given elsewhere (Sandage et al.~1996; Tammann et al.~1996;
negligible differences between these sources are due to a different
weighting of individual sources). The agreement to within the errors
between the individual luminosities supports the claim that SNe Ia are
(nearly) perfect standard candles. Independent confirmation of the
luminosities comes from H\"oflich et al.~(1996) who have three SNe Ia
in common with Table~5.  Their {\it model}\/ luminosities are the same
to within $0.12\pm0.21$ mag.  Branch's (1996) model luminosity of SN
1981B agrees fortuitously well with ours, and two SNe Ia of
Ruiz-Lapuente (1996) are fainter by only $0.26\pm 0.25$ mag judging
from their late spectra and the inferred $^{\rm 56}$Ni mass.

\begin{figure}[t]
\begin{center}
\leavevmode
\epsfxsize 12.8cm
\epsffile{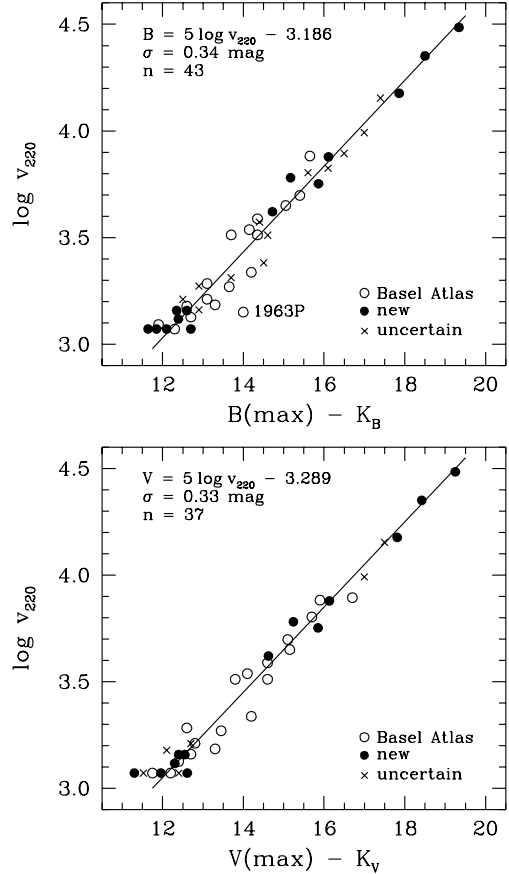}
\caption{Hubble diagrams in $B$ and $V$ of all non-red SNe~Ia with 
known maximum magnitudes. Open circles and crosses are from the 
older archive literature. Filled circles are the modern data provided by
Phillips (1993) and Hamuy et al.~(1995). The very small $K$-corrections
are applied.}
\end{center}
\end{figure}
\begin{table}[htb]
\caption{Absolute Magnitudes of SNe Ia at Maximum.}
\begin{center} \scriptsize
\begin{tabular}{lccc}
\tableline
\noalign{\smallskip}
Supernova & $M_B(max)$ & $M_V(max)$ & Reference$^a$ \\
\noalign{\smallskip}
\tableline
\noalign{\smallskip}
SN 1937C & $-19.53 \pm 0.15$ & $ -19.50 \pm 0.17$ & 1 \\
SN 1895B & $-19.87 \pm 0.22$ & --- & 2  \\
SN 1972E & $-19.52 \pm 0.22$ & $ -19.49 \pm 0.14$ & 2 \\
SN 1981B & $-19.47 \pm 0.17$ & $ -19.45 \pm 0.14$ & 3 \\
SN 1960F & $-19.53 \pm 0.14$ & $ -19.62 \pm 0.18$ & 4 \\
SN 1990N & $-19.30 \pm 0.24$ & $ -19.39 \pm 0.24$ & 5 \\
SN 1989B & $-19.51 \pm 0.21$ & $ -19.49 \pm 0.20$ & 6 \\
\noalign{\medskip}
unweighted mean & $-19.53 \pm 0.07 $ & $ -19.49 \pm 0.03$ \\
weighted mean   & $-19.53 \pm 0.07 $ & $ -19.49 \pm 0.07$ \\
\noalign{\smallskip}
\tableline\tableline
\noalign{\smallskip}
\multicolumn{4}{l}{$^a$References.-- (1) Sandage et al.~1992;  Saha et
al.~1994;} \\
\multicolumn{4}{l}{(2) Sandage et al.~1994; Saha et al.~1995; (3) Saha et
al.~1996a; } \\
\multicolumn{4}{l}{(4) Saha et al.~1996b; (5) Sandage et al.~1996; (6) Tanvir
et al.~1995\,.}\\
\end{tabular}
\end{center}
\end{table}
The Hubble diagram of {\it all\/} SNe Ia beyond 1100 km\,s$^{-1}$ with
reasonably well determined maximum magnitudes is shown in Fig.~1
(Tammann \& Sandage 1995). Their {\it intrinsic} luminosity scatter
must be considerably less than 0.35 mag, because much of the scatter
is expected from observational errors and peculiar motions. Indeed the
intrinsic scatter must be very small because even the most distant SNe
Ia lie very close to the theoretical Hubble line of slope 0.2. The
argument goes as follows. The most distant SNe Ia occupy a volume
about 18\,000 times larger than that of the local calibrators. The
large volume must contain exceptionally luminous SNe Ia -- if they
existed -- and they have a much enhanced discovery chance for two
reasons: their apparent magnitude is brighter than average {\it and}
they stay longer above the detection limit. But still, there are no
objects significantly above the Hubble line, not even at large
distances. This means: The sample of SNe Ia shown in Fig.~1
constitutes a homogeneous class of very luminous and unabsorbed
objects.

When in the following the calibration of Table 3 is applied to the SNe
Ia in Fig.~1, it should be kept in mind that \lq\lq Branch
normal\rq\rq ~SNe Ia (cf.~Branch et al.~1993) are compared with the
most luminous SNe Ia known.  Therefore the resulting value of $H_{0}$
can only be, if anything, an upper limit.

Forcing a slope of 5 (corresponding to {\it linear} expansion) to the
data in Fig.~1 gives
\begin{equation}
   B({\rm max}) = 5\,\log v -(3.186\pm 0.054)\,,
\end{equation}
and 
\begin{equation}
   V({\rm max}) = 5\,\log v -(3.289\pm 0.055)\,. 
\end{equation}
An easy calculation shows that the constant term $C_{\lambda}$
in equations (3) and (4) is determined by
\begin{equation}
   C_{\lambda}= 5\,\log H_0 - M_{\lambda} -25\,.
\end{equation}
Inserting $M_{\rm B}$ and $M_{\rm V}$ from Table~5 leads directly to
$H_0 (B)=54\pm 3$ and $H_0 (V)=58\pm 3$, from which we
adopt
\begin{equation}
   H_0= 56\pm 3 \ {\rm (internal\ error)}.
\end{equation}
Equations (3) and (4) are defined out to $30\,000$ km\,s$^{-1}$.
The value of $H_0$ therefore represents the truly cosmic expansion rate.

At a time when only the very first calibrating SNe Ia were known it was 
suggested that SN 1972E was overluminous on the basis of its light curve 
shape (Hamuy et al.~1995; Riess et al.~1995) and that consequently the true
value of $H_0$ was larger. In the light of seven calibrators this 
argument is now impossible. From first principles of stellar statistics
it is known that {\it seven nearby} objects can on average not be more
luminous than a distant, {\it luminosity-segregated} sample.
\begin{figure}[htb]
\begin{center}
\leavevmode
\epsfxsize 10cm
\epsffile{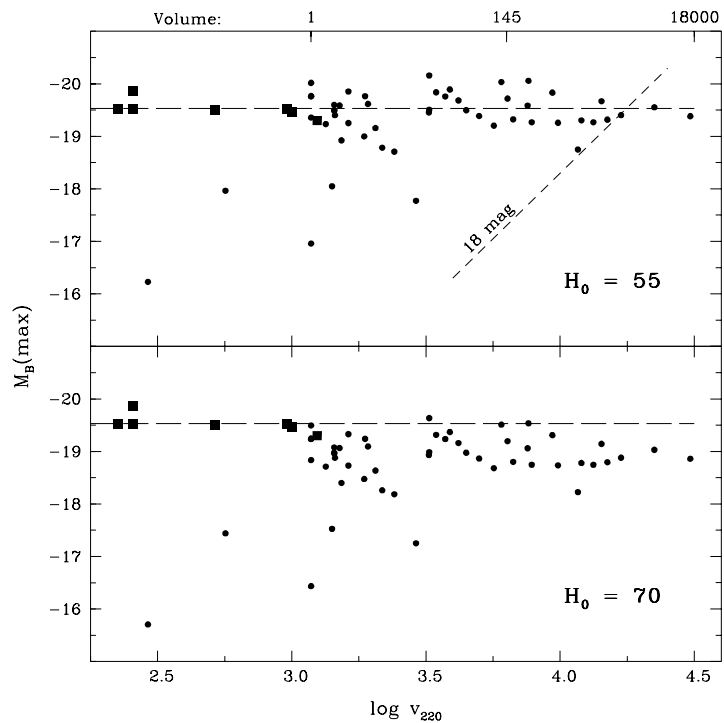}
\caption{The absolute magnitude $M_B(max)$ of all SNe~Ia in or beyond
the Virgo cluster with known $B(max)$ versus velocity distance. Also
shown are all faint {\it red} SNe~Ia; they illustrate our point that
no underluminous (or absorbed) SNe~Ia are found at large distances.
The distant objects must therefore be among the very brightest
ones. $M_B(max)$ of the calibrators (squares) is based on their
Cepheid distances. For all other SNe~Ia $M_B(max)$ is calculated from
the recession velocities and $H_0=55$ (upper panel) and $H_0=70$
(lower panel). Note that for $H_0=70$ the impossible case arises
that the distant SNe~Ia are on average less luminous than the nearby
calibrators.}
\end{center}
\end{figure}
This point is illustrated in Fig.~2, where the absolute magnitudes of
the seven calibrators are compared to the absolute magnitudes of the
distant SN sample. The latter are calculated once with $H_0=50$ and
once with $H_0=70$.  For $H_0=70$ the absurd situation arises that the
distant SN Ia are systematically fainter than the nearby
calibrators. The firm conclusion from this is that $H_0<70$\,.

A more detailed discussion of all external errors is given elsewhere
(Tammann et al.~1996). It yields a confidence range of $44<H_0<64$\,.
\section{Conclusion}
The two independent routes towards the large-scale value of $H_{0}$,
via the Virgo cluster and SNe Ia, give 54$\pm 2$ and 56$\pm 3$
(internal errors), respectively.  Their only inter\-depend\-ence is that
they rely on Cepheids (predominantly observed with {\it HST}), which are
the least controversial distance indicators at present. Together they
make a strong case for $H_{0}=55\pm 10$ (external error). Values of
$H_{0}<40$ are equally unlikely as values of $H_{0}>70$\,.

The relatively low value of $H_{0}$ is supported by additional
methods, e.g.~TF and other distances of field galaxies (Sandage 1994,
1996 and references therein), and the Zeldovich-Sunyaev effect
(Lasenby \& Hancock 1995, Rephaeli 1995).  Baum's et al.~(1995) {\it HST}
photometry of globular clusters in the Coma cluster requires
$H_{0}<65$\,. A gravitationally lensed quasar sets $H_{0}<70$ (Dahle
et al.~1994). Models of SNe Ia could not be understood if $H_{0}$ was
$\geq$ 60 (Branch et al.~1996) or in no case $\geq$70 (H\"oflich \&
Khokhlov 1996; Ruiz-Lapuente 1996).

We believe that literature values significantly larger than $H_{0}=65$
are explained by an unwarrantedly high Virgo velocity, the unrealistic
hope to fathom the depth of the Virgo cluster with only a single
galaxy, the myth of a sharp, dispersionless cutoff of the luminosity
function of planetary nebula shells, the reliance on the suspicious
surface brightness fluctuation method, and/or simply by Malmquist bias
which always artificially increases the value of $H_{0}$\,.
\acknowledgments
The Basel group acknowledges the support of the Swiss National Science
Foundation. A.S. and A.S. thank NASA for its support by grants for
publication and other expenses concerned with the overall $H_0$
project. N.P. and F.D.M. acknowledge the support of ESA as part of its
ongoing commitment to the science done with {\it HST}. We are all
grateful to the people behind the scenes of {\it HST}, without whom
none of the observations used here would have been obtained.


\begin{references}
\reference Baum, W. A., et al.~1995, \aj, 110, 2537
\reference Binggeli, B., Popescu, C. C., \& Tammann, G. A. 1993, \aaps, 98, 275 
\reference Binggeli, B., Sandage, A., \& Tammann, G. A. 1985, \aj, 90, 1681 
\reference B\"ohringer, H., et al.~1994, Nature, 368, 828
\reference Bottinelli, L., Gougenheim, L., Paturel, G., \& Teerkorpi, P.
   1991, \aap, 252, 560
\reference Branch, D., Fisher, A., \& Nugent, P. 1993, \aj, 106, 2383
\reference Branch, D., Nugent, P., \& Fisher, A. 1996, in Thermonuclear 
   Supernovae, eds. R. Canal, P. Ruiz-Lapuente, \& J. Isern (Dordrecht: 
   Kluwer Academic Publishers), in press
\reference Burstein, D. \& Heiles, C. 1984, \apjs, 54, 33
\reference Capaccioli, M., Della Valle, M., D'Onofrio, M., \& Rosino, L. A.
   1989, \aj, 97, 1622
\reference Ciardullo, R., Jacoby, G. H., \& Ford, H. C. 1989, \apj, 344, 715
\reference Chiosi, C., Wood, P., \& Capitanio, N. 1993, \apjs 86, 541
\reference Cohen, J. G. 1985, \apj, 292, 90
\reference Dahle, H., Maddox, S. J., \& Lilje, P. B. 1994, \apjl, 435, L79
\reference Dressler, A. 1987, \apj, 317, 1
\reference Elson, R. A. W., \& Santiago, B. X. 1996, \mnras, in press
\reference Faber, S. M., et al.~1989, \apjs, 69, 763
\reference Feast, M., \& Walker, A. R. 1987, \araa, 25, 345
\reference Federspiel, M., Tammann, G. A., \& Sandage, A. 1996, in press.
\reference Freedman, W. L. \& Madore, B. F. 1990, \apj, 365, 186
\reference Freedman, W. L. \& Madore, B. F. 1994, \apj, 427, 628
\reference Freedman, W. L., et al. 1994, Nature 371, 757
\reference Fritze-von Alvensleben, U. 1995, private communication
\reference Gould, A. 1994, \apj, 425, 51
\reference Hamuy, M., et al.~1995, \aj, 109, 1
\reference Harris, W. E. 1990, \pasp, 102, 966
\reference H\"oflich, P., \& Khokhlov, A. 1996, \apj, 457, 500
\reference H\"oflich, P., \& Khokhlov, A., Wheeler, J.C., Nomoto, K., \& 
   Thielemann, F.K. 1996, in Thermonuclear Supernovae, eds. R. Canal, 
   P. Ruiz-Lapuente, \& J. Isern (Dordrecht: Kluwer Academic Publishers), 
   in press
\reference Huchra, J. 1988, in The Extragalactic Distance Scale, eds. S. van 
   den Bergh \& C. J. Pritchet, (San Francisco: Astronomical Society
   of the Pacific), 257
\reference Jacoby, G. H., Ciardullo, R. 1993, in Planetary Nebulae, 
   eds. R. Weinberger \& A. Acker, (=IAU Symposium 155), 503 
\reference Jacoby, G. H., Ciardullo, R., Booth, J., \& Ford, H. C., 1989,
   \apj, 344, 704
\reference Jacoby, G. H., Ciardullo, R., \& Ford, H. C. 1990, \apj, 356, 332
\reference Jerjen, H. 1995, private communication
\reference Jerjen, H., \& Tammann, G. A. 1993, \aap, 276, 1
\reference Kelson, D. 1995, private communication
\reference Kennicutt, R. C., Freedman, W. L., \& Mould, J. R. 1995, 
   \aj, 110, 1476
\reference Kraan-Korteweg, R. C. 1986, \aaps, 66, 255
\reference Lasenby, A. N., \& Hancock, S. 1995, in Current Topics in 
   Astrofundamental Physics: The Early Universe, eds. N. Sanchez \& 
   A. Zichichi, (Dordrecht: Kluwer Academic Publishers), 327
\reference Lorenz, H., B\"ohm, P., Capaccioli, M., Richter, G. M., \& Longo,
   G. 1993, \aap, 277, L15
\reference Madore, B.F., \& Freedman, W. L. 1991, \pasp, 103, 933
\reference M\'endez, R. H., Kudritzki, R. P., Ciardullo, R., \& Jacoby, G. H.
   1993, \aap, 275, 534
\reference Mould, J. R., et al.~1995, \apj, 449, 413
\reference Panagia, N., Gilmozzi, R., Macchetto, F. D., Adorf, H.-M., \&
   Kirshner, R. P. 1991, \apjl, 380, L23
\reference Phillips, M. M. 1993, \apjl, 413, L105
\reference Phillips, M. M., Jacoby, G. H., Walker, A. R., Tonry, J. L., \&
   Ciardullo, R. 1992, BAAS, 24, 749
\reference Pritchet, C. J., \& van den Bergh, S. 1987, \apj, 318, 507
\reference Rephaeli, Y. 1995, \araa, 33, 541
\reference Riess, A. G., Press, W. H., \& Kirshner, R. P. 1995, \apjl, 438, L17
\reference Ruiz-Lapuente, P. 1996, preprint
\reference Saha, A., Labhardt, L., Schwengeler, H., Macchetto, F. D., 
   Panagia, N., Sandage, A., \& Tammann, G. A. 1994, \apj, 425, 14 (IC 4182)
\reference Saha, A., Sandage, A., Labhardt, L., Schwengeler, H., 
   Tammann, G. A., Panagia, N., \& Macchetto, F. D. 1995, \apj, 438, 8 (NGC 
5253)
\reference Saha, A., Sandage, A., Labhardt, L., Tammann, G. A., 
   Macchetto, F. D., \& Panagia, N. 1996a, \apj, in press (NGC 4536)
\reference Saha, A., Sandage, A., Labhardt, L., Tammann, G. A., 
   Macchetto, F. D., \& Panagia, N. 1996b, \apj, in press (NGC 4496A)
\reference Sandage, A. 1988, \pasp, 100, 935 
\reference Sandage, A. 1994, \apj, 430, 1
\reference Sandage, A. 1996, \aj, 111, 1 and 18
\reference Sandage, A., \& Bedke, J. 1994, Atlas of Galaxies (Washington, D.C.:
   NASA)
\reference Sandage, A., Saha, A., Tammann, G. A., Labhardt, L., Panagia, N., 
   \& Macchetto, F. D. 1996, \apjl, in press
\reference Sandage, A., Saha, A., Tammann, G. A., Labhardt, L., 
   Schwengeler, H., Panagia, N., \& Macchetto, F. D. 1994, \apjl, 423, L13 
   (NGC 5253)
\reference Sandage, A., Saha, A., Tammann, G. A., Panagia, N., \& 
   Macchetto, F. D. 1992, \apjl, 401, L7 (IC 4182)
\reference Sandage, A., \& Tammann, G. A. 1968, \apj, 151, 531
\reference Sandage, A., \& Tammann, G. A. 1974, \apj, 194, 223
\reference Sandage, A., \& Tammann, G. A. 1990, \apj, 365, 1
\reference Sandage, A., \& Tammann, G. A. 1995, \apj, 446, 1
\reference Sandage, A., \& Tammann, G. A. 1996, \apjl, in press
\reference Schaefer, B.E. 1995, \apjl, 447, L13
\reference Schr\"oder, A. C. 1995, Ph.D. Thesis, Univ. Basel
\reference Soffner, T., et al. 1995, \aap, in press
\reference Tanvir, N. R., Shanks, T., Ferguson, H. C., \& Robinson, D. R. T. 
1995, Nature, 377, 27
\reference Tammann, G. A. 1988, in The Extragalactic Distance Scale, eds. 
   S. van den Bergh \& C. J. Pritchet, (San Francisco: Astronomical Society
   of the Pacific), 282
\reference Tammann, G. A. 1993, in Planetary Nebulae, eds. R. Weinberger \& 
   A. Acker, (=IAU Symposium 155), 515
\reference Tammann, G. A. 1996, Rev. Modern Astronomy, in press
\reference Tammann, G. A., \& Sandage, A. 1968, \apj, 151, 825 
\reference Tammann, G. A., \& Sandage, A. 1995, \apj, 452, 16 
\reference Tammann, G. A., Sandage, A., Saha, A., Labhardt, L., 
   Macchetto, F. D., \& Panagia, N. 1996, in Thermonuclear Supernovae, 
   eds. R. Canal, P. Ruiz-Lapuente, \& J. Isern (Dordrecht: Kluwer 
   Academic Publishers), in press  
\reference Tonry, J. L. 1991, \apjl, 373, L1
\reference Tonry, J. L., Ajhar, E. A., \& Luppino, G. A. 1990, \aj, 100,1416
\reference Tonry, J. L., \& Schneider, D. P. 1988, \aj, 96, 807
\reference Whitmore, B. C., et al. 1995, \apj, 454, 773
\end{references}
\end{document}